\documentclass[a4paper,11pt,nofootinbib,aps,superscriptaddress]{article}
\usepackage[utf8]{inputenc}
\usepackage{graphicx}
\usepackage{hyperref}      
\usepackage[usenames,dvipsnames]{color} 
\usepackage{amssymb}
\usepackage{amsmath}
\usepackage{titlesec}

\usepackage[sort&compress,numbers]{natbib}

\usepackage{authblk}

\usepackage{geometry}
\newgeometry{vmargin={24mm}, hmargin={26mm,26mm}}

\def\lsim{\raise-.75ex\hbox{$\buildrel<\over\sim$}}

\begin{document}

{\huge \begin{center} {\bf ESA Voyage-2050 White Paper}  \end{center}}

 {\Large \begin{center} {\bf Mapping Large-Scale-Structure Evolution over Cosmic Times} \end{center}}

\begin{figure}[h!]
\centering
\includegraphics[width=\textwidth]{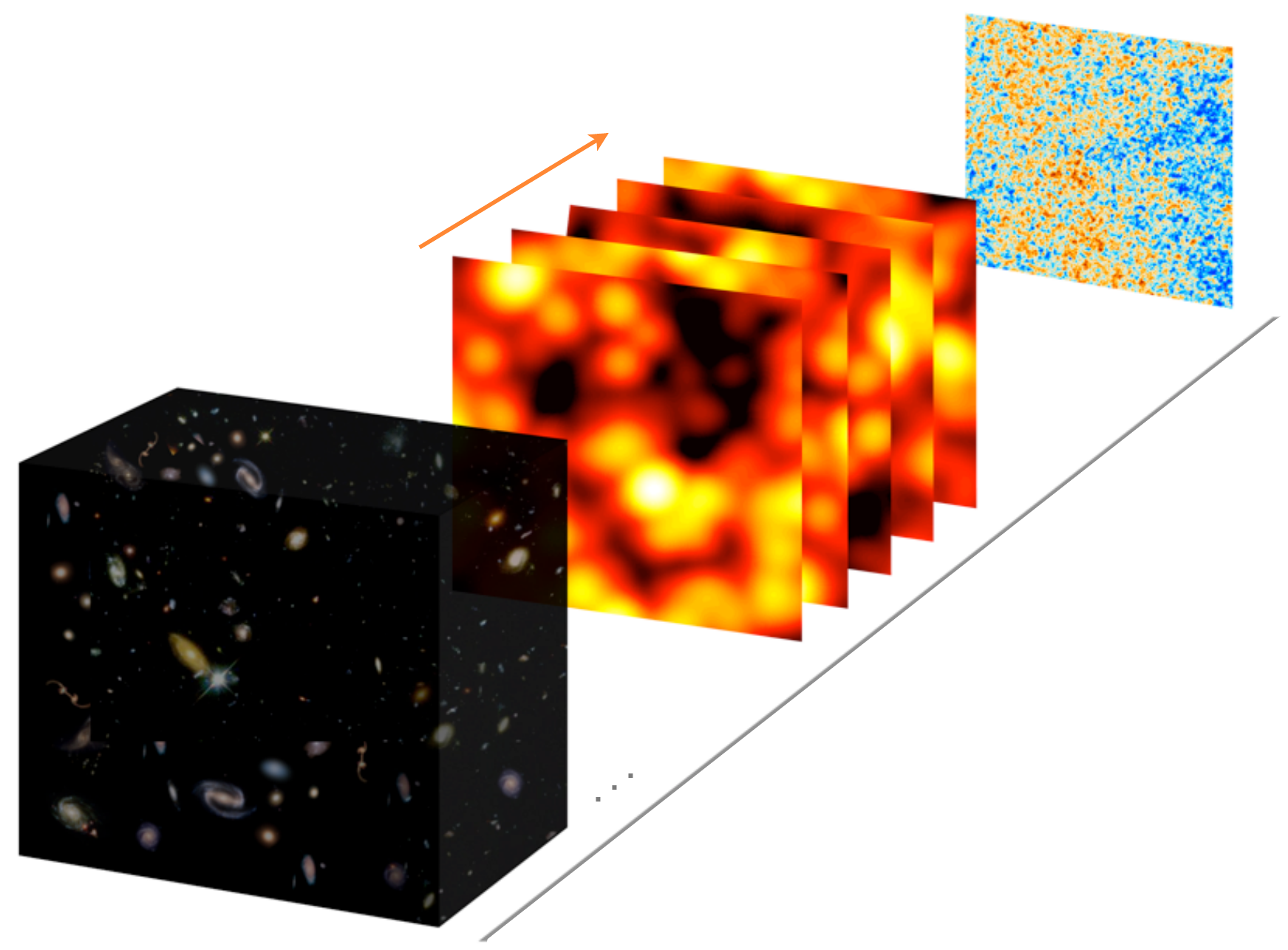}
\label{fig:IMVisual00}
\end{figure}

\vspace{0.3in}

\begin{center}
Principal Author: 
\textnormal{Marta B.~Silva}
\\
\textnormal{Institution: Institute of Theoretical Astrophysics, University of Oslo}
\\
\textnormal{Email: m.b.silva@astro.uio.no}
\\
\textnormal{Phone: +47 22 857 632}
\\
\textnormal{Address: Sem Sælands vei 13, Svein Rosselands hus, 0371 Oslo, Norway
}
\end{center}

\newpage

\author[1]{Marta B.~Silva}
\author[2]{Ely D. Kovetz}
\author[3]{Garrett K. Keating}
\author[4,5]{Azadeh Moradinezhad Dizgah}
\author[6]{Matthieu Bethermin}
\author[7]{Patrick C. Breysse}
\author[8]{Kirit Kartare}
\author[9]{Jos\'e L. Bernal}
\author[10,11]{Jacques Delabrouille}

\affil[1]{ Institute of Theoretical Astrophysics, University of Oslo, P.O.B 1029 Blindern, N-0315 Oslo, Norway}
\affil[2]{Department of Physics, Ben-Gurion University, Be'er Sheva 84105, Israel}
\affil[3]{Harvard-Smithsonian Center for Astrophysics, 60 Garden Street, Cambridge, MA 02138, USA}
\affil[4]{Departement of Theoretical Physics and Centre for Astroparticle Physics (CAP), University of Geneva, 24 quai E. Ansermet, CH-1211 Geneva, Switzerland}
\affil[5]{Department of Physics, Harvard University, 17 Oxford St., Cambridge, MA 02138, USA}
\affil[6]{Aix Marseille Universit\'e, CNRS, CNES, LAM, Marseille, France}
\affil[7]{Canadian Institute for Theoretical Astrophysics, University of Toronto, Ontario, M5S 3H8, Canada}
\affil[8]{Kavli Institute for Cosmological Physics, University of Chicago, Chicago, IL 60637, USA}
\affil[9]{ICC, University of Barcelona, IEEC-UB, Mart\'ı i Franques 1, E08028 Barcelona, Spain}
\affil[10]{Laboratoire Astroparticule et Cosmologie, CNRS/IN2P3, 75205 Paris Cedex 13, France}
\affil[11]{D\'epartement d’Astrophysique, CEA Saclay DSM/Irfu, 91191 Gif-sur-Yvette, France}

\date{}                     
\setcounter{Maxaffil}{0}
\renewcommand\Affilfont{\itshape\small}

\title{ ESA Voyage-2050 White Paper: \\  Mapping Large-Scale-Structure Evolution over Cosmic Times}

\maketitle

\vspace{-0.4in}

\tableofcontents

\vspace{0.4in}


\begin{abstract}

\vspace{0.1in}

This paper outlines the science case for line-intensity mapping with a space-borne instrument targeting the sub-millimeter (microwaves) to the far-infrared (FIR) wavelength range. 
Our goal is to observe and characterize the large-scale structure in the Universe from present times to the high redshift Epoch of Reionization. 
This is essential to constrain the cosmology of our Universe and form a better understanding of various mechanisms that drive galaxy formation and evolution. We argue that the proposed frequency range would make it possible to probe important metal cooling lines such as [CII] up to very high redshift as well as a large number of rotational lines of the CO molecule. These can be used to trace molecular gas and dust evolution and constrain the buildup in both the cosmic star formation rate density and the cosmic infrared background (CIB).  Moreover, surveys at the highest frequencies will detect FIR lines which are used as diagnostics of galaxies and AGN. Tomography of these lines over a wide redshift range will enable invaluable measurements of the cosmic expansion history at epochs inaccessible to other methods, competitive constraints on the parameters of the standard model of cosmology, and numerous tests of dark matter, dark energy, modified gravity and inflation. 
To reach these goals, large-scale structure must be mapped over a wide range in frequency to trace its time evolution over a reasonable fraction of the volume of the observable Universe. In addition, the surveyed area needs to be very large to beat cosmic variance and to probe the largest scales where its easier to separate the astrophysical and cosmological contributions to the observed signal. Only, a space-borne mission can properly meet these requirements. 

\end{abstract}

\eject
\pagenumbering{arabic} 
\setcounter{page}{1}

\section{Introduction}

\subsection{The promise of line-intensity mapping}

Line-intensity mapping (LIM) \cite{Kovetz:2017agg} is an emerging technique to explore galaxy and structure evolution over cosmic times by collecting all incoming photons along the line of sight at a given frequency, whether they originate from within galaxies or the intergalatic medium, and measuring the spatial fluctuations in the emission. The fluctuation maps provide a tracer of both the underlying density fluctuations, which are affected by the detailed makeup and evolution of the cosmic inventory, and by the physical processes that govern the radiation sources. Access to the full sky, with only modest angular resolution requirements---ideal terms for space-borne instruments---allows a global investigation of different epochs in the history of the Universe (from cosmic dawn and reionization to modern times) and of various processes such as galaxy formation and evolution, star-formation history, metal and dust buildup in galaxies, (proto)cluster properties, identifying the extragalactic background light sources, etc.

Maps of line-intensity fluctuations are uniquely advantageous over those of cosmic microwave background fluctuations or of pixelised number counts of discrete galaxy surveys. Compared to the former, LIM is not limited by diffusion damping on small scales and can be measured in tomography over huge cosmological volumes, which unlike the latter, LIM is not bound to a census of discrete bright sources and can extend to very high redshifts. Figure 1 (adapted from Refs.~\cite{Kovetz:2017agg,Kovetz:2019uss}) illustrates the power of LIM.

\begin{figure}[h!]
\centering
\includegraphics[width=\textwidth]{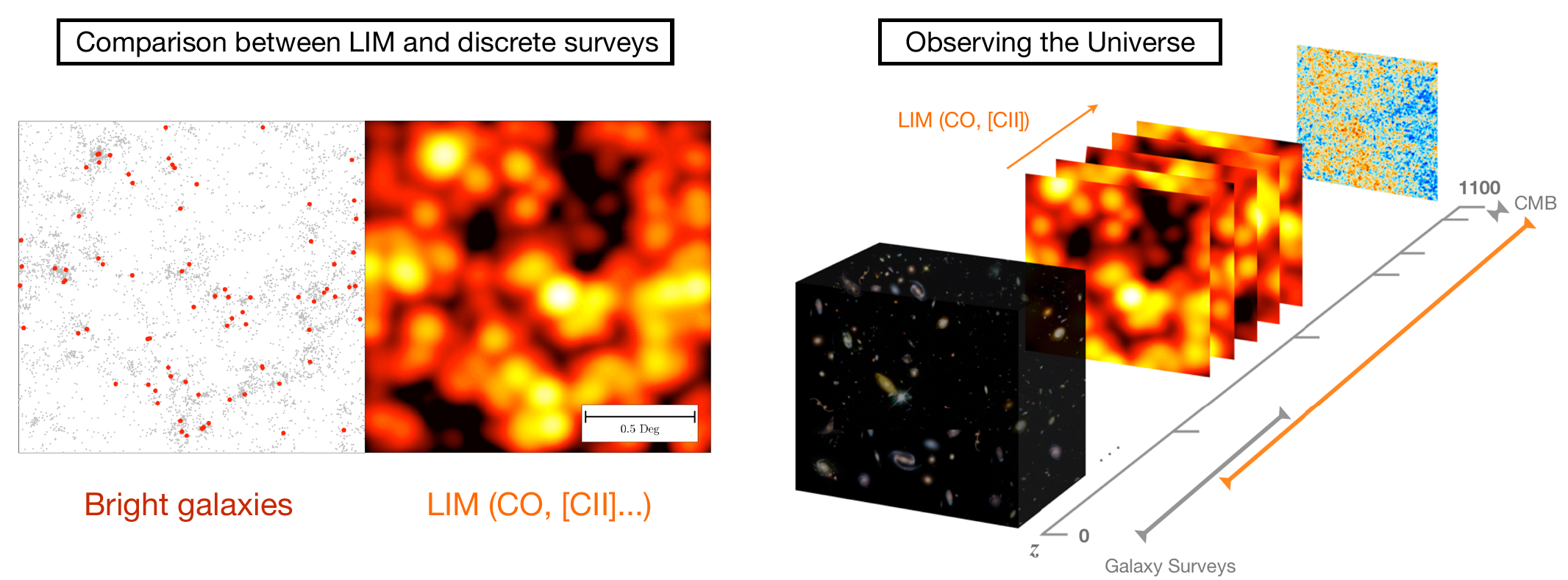}
\caption{{\it Left:} A comparison between the measurements of discrete bright galaxies by the VLA and the fluctuations in the total CO emission from all the galaxies by the COMAP experiment, if similar observation time is dedicated by both to a small patch of sky. {\it Right:} LIM can probe the $\gtrsim\!80\%$ volume of the observable Universe which is out of reach of CMB and galaxy surveys.}
\label{fig:IMVisual0}
\end{figure}

Targets for LIM cover a large swath of the electromagnetic spectrum, from the 21-cm emission from neutral hydrogen in the intergalactic medium from the cosmic dawn of star formation ($z\, \lesssim\, 25$), to Lyman-$\alpha$ from young star forming galaxies at the epoch of reionization (EoR) ($z\, \gtrsim\, 6$) and  epoch of cosmic buildup ($2\, \lesssim\, z\, \lesssim 6$). 
In between lies is a plethora of HI, He and metal  galaxy lines tracing structures over cosmic times, ranging from the EoR down to nearby times ($z\, \sim 0.1$, close to one billion years ago), which provide a varied and complementary set of powerful probes of astrophysics and cosmology. A sensitive space mission that can access several of these lines over a wide range of frequencies will address an extensive list of science goals, as described below.

\subsection{Intensity Mapping at Millimeter and Far Infrared Wavelengths}
\label{Sec:1.2}
\begin{figure}[h!]
\centering
\includegraphics[width=\linewidth]{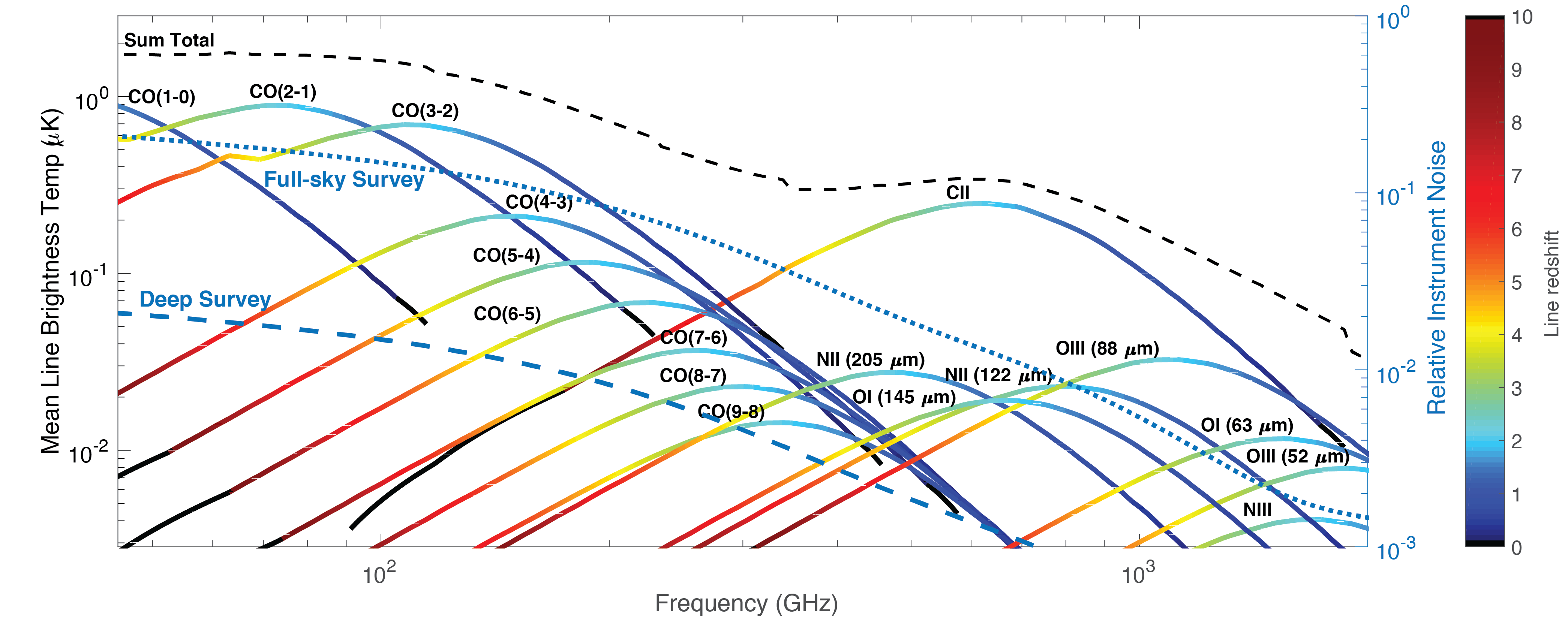}
\caption{ Spectral lines detectable by a sky survey in submillimiter to high-frequency radio waves. Shown above are model predictions for the line intensities and the relative sensitivities of the full-sky and "deep" ($400$ deg$^2$) surveys. The full survey will be used to perform statistical tomography measurements and constrain cosmological models. The deep survey will enable direct imaging of the emission sources, cross-correlations with other surveys, precise foreground subtraction and pinpointing the localization of galaxy (proto)clusters well into the EoR. The shown signal strengths adopt line luminosities scaled from the IR luminosity based on observational relations. Current constraints on these line ratios are uncertain by up to one order of magnitude.}
\label{fig:IMVisual}
\end{figure}

The millimeter, sub-millimeter, and far infrared  regimes offer a number of natural targets for LIM, spanning a wide range of redshifts, with the potential for probing the multitude of galaxies stretching back to the EoR. 

As shown in Figure \ref{fig:IMVisual}, the window between ${\rm 1\, cm}$ (${\rm 30\, GHz}$) and ${\rm 100\, \mu m\, (3\, THz)}$ contains several bright rotational and fine-structure lines. The several line strengths were modelled as a function of IR luminosity using observationally-based scaling relations \citep{Sargent2012, Bonato2019}. These measurements include local and high-z galaxies and are assumed to be constant with redshift. The IR luminosities used to compute the line intensities, and bias were derived from the SFR in the Eagle (Evolution and Assembly of GaLaxies and their Environments) simulation \cite{Schaye2015}, which constrains these relations as a function of redshift. The SFRD derived from the Eagle simulation shows a good fit to observational constraints based on UV band measurements. Due to the lack of additional constraints, the modelling of these lines is uncertain by a factor of a few (low-z) to an order of magnitude towards high-z ($z\, >\, 6$). The modelling of the CO/[CII] lines based on other analytical calculations and galaxy simulations face a similar large uncertainty due to a lack of constraints on a large number of free parameters \citep[see e.g.]{Mashian2015}). These models can only be meaningfully improved with a high sensitive LIM space mission over a broad frequency range to trace the redshift evolution of the CO/[CII] line strengths up to high-z. Note that the possible CMB effect on suppressing the low CO line transitions and enhancing higher order transitions at high-z was not accounted for in this modelling; this effect is due to the CMB being more efficient at heating the gas during the EoR \citep[see e.g.]{Vallini2018}. Further corrections due to the CO lines being observed against the CMB are dependent on how the CMB is removed from the observations and are mostly a problem in the context of individual galaxy surveys \citep[see e.g.]{daCunha2013}.

The considered target lines are not easily accessible from ground-based instruments, as the atmosphere is only transparent over a small fraction of the quoted window. However, a space-based instrument would offer the potential of increased sensitivity and continuous spectral coverage over this entire spectral range. This continuous coverage is particularly valuable for LIM studies, as it affords the potential for cross-correlation between lines, as well as better access to the largest spatial scales for studying the structure of the Universe.

\begin{figure}[h!]
\centering
\includegraphics[width=\linewidth]{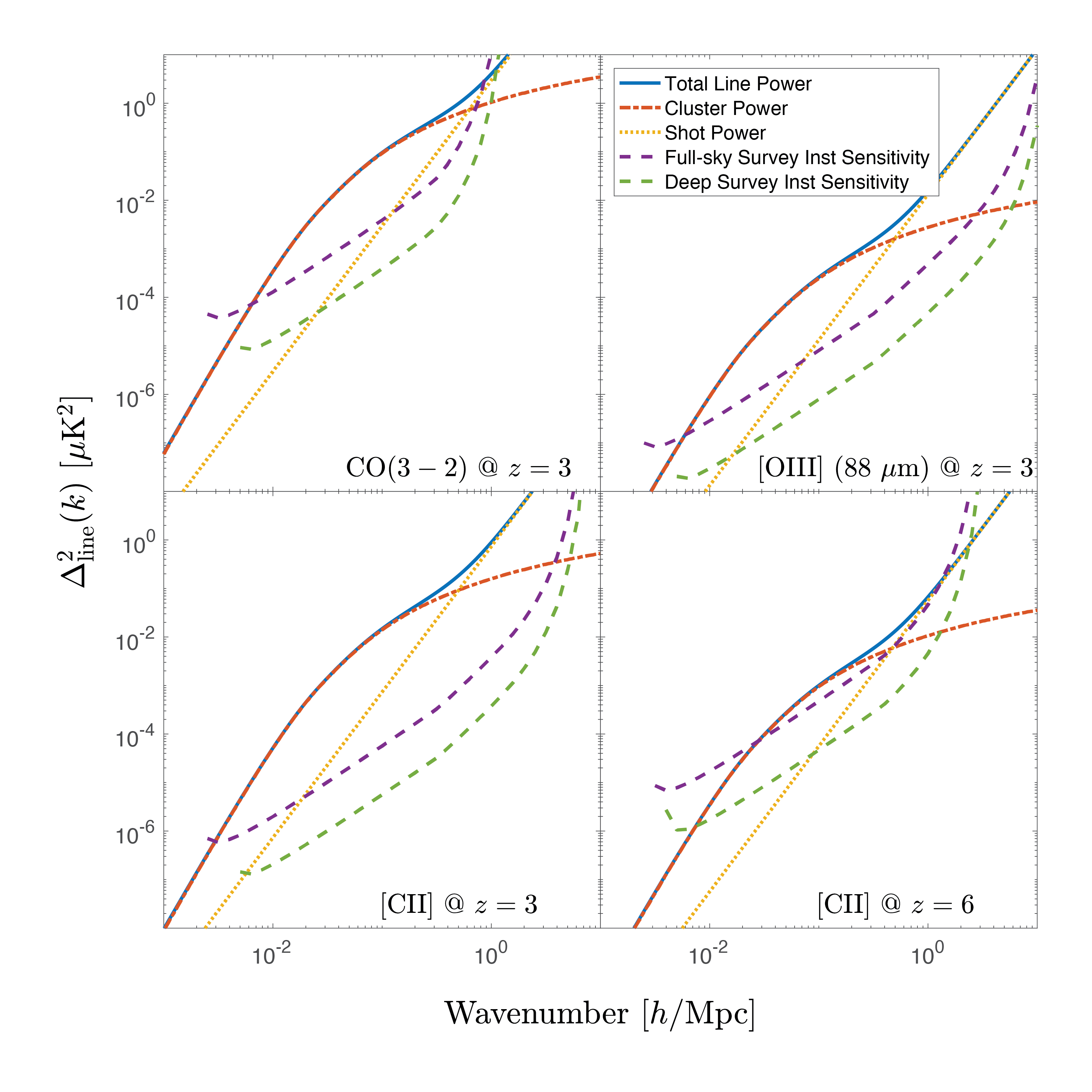}
\vspace{-1.0 cm}
\caption{Estimated instrument sensitivities for the deep and full-sky surveys (purple dashed and green dash, respectively), versus models predictions for several lines of interest, with the cluster (orange dot-dash), shot (yellow dotted), and total power (blue) shown separately. A full-sky survey with a space-based instrument would be expected to place tight power spectrum constraints on CO and [CII] emission at $z\approx3$ (left panels), over a large range of spatial scales from $k\approx0.01-10$ h/Mpc. A deeper survey on a smaller region of sky (i.e., $f_{\rm sky}=0.01$) could achieve a high signal-to-noise detection of the [CII] EoR power spectrum (lower left panel), as well as several fainter line species like the higher-J ($J_{\rm upper} \geq 6$) transitions of CO. Such an instrument would also be capable of measuring the [OIII] 88 $\mu$m line (upper right).}
\label{fig:IMPowSpec}
\end{figure}

This paper describes the various science goals achievable with a space mission spanning a frequency range of up to 50-2000 GHz. As shown in Figure \ref{fig:IMVisual}, such a potential mission would be particularly capable of LIM analyses using the rotational transitions of CO, as well as the [CII] 158 $\mu$m line; it would make it possible to probe the EoR with [CII] by leveraging the broad spectral coverage to remove confusion and contamination of the higher rotational lines of CO, to a level that is well beyond the access of discrete galaxy surveys. Moreover, it would make spectrally rich measurements during the peak of cosmic star formation at $z\sim2$, using [NII], [OI], and [OIII] as additional probes of a broad population of galaxies.

For this hypothetical space mission, we assume an aperture size of 3.5m, actively cooled to 8 K with 1\% emissivity, with a medium-resolution $R=300$ spectral camera, with 64 independent dual-polarization beams, and total optical efficiency of 25\%. Shown in Figure \ref{fig:IMPowSpec} are the expected power spectrum sensitivities of two different $\tau=10^4$ hour surveys (running order 1-2 years in total observing length). A full-sky survey would be capable of making high-significance power spectrum detections of both CO and [CII] over a broad range of redshifts, while a deeper survey focusing in on only 1$\%$ of the sky ($\approx400$ sq. deg.) would be capable of making deep power spectrum measurements of [CII] in the EoR, as well as of the higher rotational line species of CO (e.g., CO(6-7), CO(8-7), CO(9-8)) and the 122 $\mu$m and 205 $\mu$m lines of [NII].

\subsubsection{[CII] line intensity mapping}\label{ssec:cii_mapping}

The [CII] 158 $\mu$m line is a major cooling line in galaxies spectrum, and its tightly correlated with continuum infrared emission in both normal star forming galaxies and in non-AGN luminous infrared galaxies \cite{Herrera2015}. 

A [CII] LIM mission covering a large area in space and a broad range in time can, therefore, constrain the buildup of the continuum infrared background (CIB). This goal is out of the reach of traditional galaxy surveys which cannot detect large numbers of dusty galaxies at $z\gtrsim3$. [CII] maps probe galaxies emission reprocessed by dust significantly improving upon current constraints on the cosmic SFRD and pinpointing the sources of the dust-obscured star formation. Moreover, [CII] intensity maps can be used to construct more accurate sky maps to use in the processing of the CMB. Better sky maps could potentially be used to improve the accuracy of several cosmology parameters derived from CMB missions. 

The ideal [CII] LIM mission would cover a broad frequency ranging from 200 - 2000 GHz (CII at $z\sim 8-1$) with a field of view of the order of 25 - 100 deg$^2$ to beat cosmic variance.
At this frequency range, the [CII] maps will be contaminated by emission from several interloping lines, namely several CO rotational transition lines. Extending the survey frequency cover down to 50 - 100 GHz would make it possible to detect multiple CO transition lines originated from the same low redshift structures. This would yield precise information about the properties and spatial distribution of the structures responsible for the CO emission, making it possible to accurately discriminate between the signal from different lines, and to isolate the [CII] signal up to very high redshift. Additionally, 1 - 2 arcmin (or at least $\sim 5$ arcmin) angular resolution, is needed to image the individual large scale filaments responsible for the low redshift CO lines emission, making it possible to disentangle the different lines in map space with high precision. For reference, at $z\, =\, 1$, the correspondence between emitting structure size and angular resolution is such that 1$^\prime$ ($\sim$1 Mpc) and (1 - 2) deg correspond respectively to the characteristic width and length of a large-scale filament; ``small" IGM filaments have similar widths but characteristic lengths of 5$^\prime$ to 6$^\prime$. 

The proposed sky survey would revolutionize the [CII] LIM field by producing tomographic maps of this line and providing the interloping lines data necessary to clean them from contamination. Most of the existing/planned LIM missions targeting the [CII] line will only target a narrow range in redshift in the post EoR. A few missions target [CII] from the EoR but the large degree of contamination by CO imposes substantial requirements for a survey aiming at recovering clean [CII] maps at $z>6$ \cite{Sun2018}. New methods are being developed to address this line foreground issue. But in any case current missions can detect the [CII] signal in cross-correlation with other surveys, or possible make a statistical signal detection.

\subsubsection{ CO line intensity mapping}\label{ssec:co_mapping}

CO rotation lines are the most commonly used tracers of molecular gas. These lines are emitted in a ladder with a frequency separation of 115.271 GHz. The CO spectral lines energy distribution (SLED) is a strong diagnosis tool of the properties of galaxies ISM.
The conversion between CO emission and molecular gas is uncertain by a factor of a few for the CO(1-0) line and even larger for high-J CO lines~\cite{Mashian:2015his, Mashian:2016bry}. However, LIM of multiple CO lines emitted from the same structures can be used to determine the CO SLED which will constrain the state of the ISM gas. Knowledge of the ISM state can be used to infer the appropriate CO to molecular gas mass conversion factor, resulting in much tighter constraints on the ladder. A LIM mission covering the 100 - 2000 GHz frequency range will be able to detect emission from multiple CO rotation lines over a broad range in redshift. Towards higher redshifts, the CMB was warmer, and so its excitation effect on ISM gas  might have driven the CO ladder to peak at high-J transitions \cite{Vallini2018}. While this is a poorly explored field, due to a lack of observational constraints, this effect can make high-J transitions easier to observe. 

\section{Galaxy Evolution}

LIM is the ideal tool to address several open questions surrounding galaxies and their evolution: 
\begin{itemize}
    \item \emph{The cosmic star formation rate density:} What is the contribution of the faint galaxy population to the cosmic star formation rate density? What fraction of cosmic starlight is absorbed by dust and re-radiated at shorter wavelengths? 
    \item \emph{Galaxy properties:} How different was the ISM of the first galaxies, and how did it evolve with time? What are the histories of metal enrichment and dust buildup in galaxies? 
    \item \emph{The galaxy - large scale structure (LSS) connection:} How do galaxy properties vary with their large-scale environment? Do different galaxy spectral lines follow the same underlying LSS? How does galaxy clustering evolve across cosmic time? 
\end{itemize}

The unique science potential associated with LIM is the result of a combination of unbiased measurements over large enough volumes to beat cosmic variance, sensitivity to faint objects, and precise redshift measurements. An ambitious LIM space mission will provide a powerful complement to resolved galaxy observations with current and future surveys by Euclid and ALMA, or NASA  observatories such as JWST, WFIRST, HabEx, LUVOIR and OST.

\subsection{The Cosmic Star Formation history} 
\begin{figure}[h!]
\centering
\includegraphics[width=\textwidth]{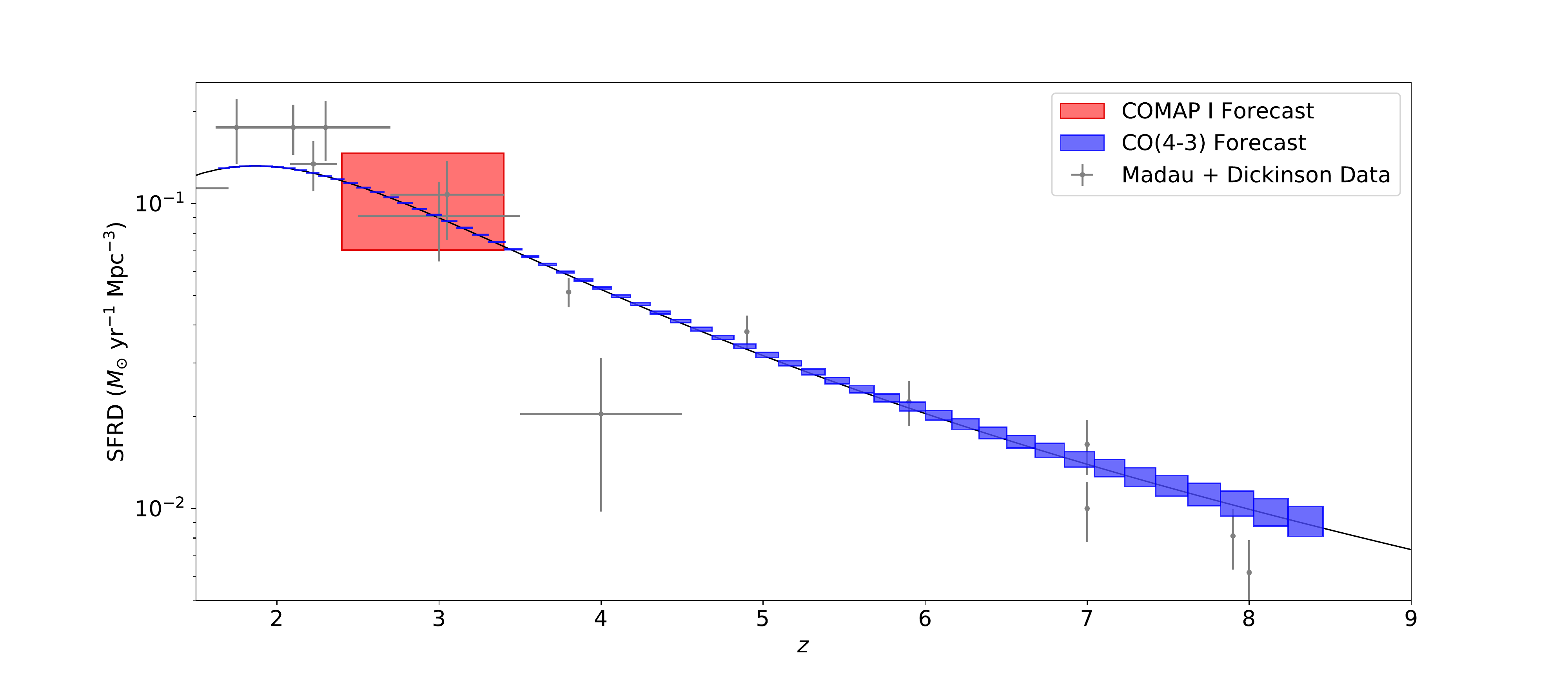}
\caption{Forcast for constraints on the SFRD across cosmic times using a combination of the one and two point functions~\cite{Breysse:2016szq,Ihle:2018dky}. The proposed space-borne LIM instrument can derive tight constraints on the SFRD with the CO(4-3) line up to $z\,=\, 8$; in this model the line ratio between CO and LIR is constant with redshift. Moreover, this instrument can derive similar constraints with the [CII] line or with higher-J CO lines.
Above $z\sim3$ the SFRD is mostly derived from rest-frame UV measurements (shown in the figure; taken from Ref.~\cite{Madau:2014bja}), the data points showing the huge uncertainty on using observations solely in this frequency band. Below $z=3$ galaxy surveys probe the dust-enshrouded star formation with metal lines, but at higher redshifts this is increasingly challenging. Recently, an unexpectedly large number of $z>3$ dusty and massive starburst galaxies was detect by the ALMA telescope \cite{2019arXiv190802372W}, these galaxies were detected in the submillimeter and are invisible in the ultraviolet to infrared. At $z>6$, for dust rich models, we can be missing more than 90$\%$ of the total SFRD \cite{Casey2018}.}
\label{fig:IMVisual2}
\end{figure}

Constraining the cosmic SFRD and understanding the main mechanisms driving its evolution is a major goal of astrophysics. 
The rate of star formation in galaxies is a function of many variables, ranging from the amount and composition of star-forming gas to complex mechanisms which affect the rate at which this gas is converted into stars. Many of these factors are challenging to access through observations, namely galaxy history and environment.  At $z\gtrsim2.5$, measurements of the SFRD are mostly derived from optical and ultraviolet luminosity functions. Of the order of half of the starlight is re-processed by dust \cite{Casey2018}, leading to a large uncertainty relative to how much star formation is being missed by observations at high-z. Moreover, from the SFRD alone, it is not possible to disentangle star-formation efficiency from the molecular gas mass available to fuel star formation in galaxies. Inflows of gas from the IGM make it possible for galaxies to continue forming stars long after the cold-gas depletion time \cite{Spring2017}. Therefore, environmental effects such as clustering and location relative to the underlying cosmic web are vital to understanding the global star formation process.
{\em LIM yields a global census of several gas phases and of stellar buildup across cosmic time.} LIM can trace the fuel for star formation using CO lines \cite{Breysse2016A}, instantaneous star formation with H-$\alpha$ and Ly$\alpha$ \citep{Silva2013, Gong2017, Silva2017} and star formation on long timescales with CII and other lines \cite{Silva:2014ira}. Intensity maps also contain information about galaxy clustering and the location of the sources relative to the underlying LSS and so they can place galactic emission in its large-scale environmental context.

\subsection{The High-Redshift ISM}

The properties of high redshift galaxies are likely to differ substantially from today's systems. The first galaxies are thought to be on average small, dense, poor in metal and dust, and hosts of young stellar populations. Due to frequent mergers, the ISM of these galaxies was probably more turbulent, leading to boosted star-formation efficiency \cite{Chaikin2018}. Also, the circumgalactic medium (CGM) temperature and ionization state were set by a much weaker extragalactic background radiation. 
In addition, due to the relatively young age of the universe, it is expected that there is a large scatter in relations between galaxy properties and line luminosity compared to the low-z universe. Therefore, probing the average properties of the first galaxies requires observing a vast number of sources. Future galaxy surveys will struggle to detect enough average main sequence star-forming galaxies to converge on their properties, especially at $z\gtrsim 8$.

\emph{LIM can access multiple spectral lines characteristic of different phases of the ISM and the CGM, thus constraining the properties of high-z galaxies.} Moreover, these are unbiased surveys over large volumes that probe emission from both bright and faint sources. 
LIM can be used to probe multiple CO rotational-transition lines which trace the distribution of gas densities and temperatures in the molecular clouds where stars form \cite{Vallini2018}. CII emission from photo-dissociation regions provides one of the most significant sources of cooling \cite{Dalgarno1972} in galaxies, and it is a reliable tracer of star formation down to low metallicity levels. HI 21-cm line emission at low-z ($z\lesssim 1$) is a useful probe of neutral gas in galaxies, and at high-z ($z\gtrsim 6$) it traces neutral gas in the IGM. Also, H-$\alpha$, and Lyman-$\alpha$ probe ionized recombining gas, therefore tracing instantaneous star formation.  Meanwhile, the Lyman-$\alpha$ line can probe several characteristics of the CGM, such as its ionization state and the gas density distribution \cite{Villaescusa-Navarro2018, Pullen:2013dir, Silva2013}, up to $z\sim10$ \citep{2017Heneka, Cooray2019}. [OIII], used by ALMA to measure redshifts of $z=7-9$ galaxies~\cite{Moriwaki2018}, can also be targeted with LIM to study the ionization state of gas in the CGM \citep{Gong2017, Silva2017}. 
Another promising means to study molecular gas density and composition is the use of cross correlation between CO isotopologues~\cite{2017MNRAS.468..741B,Zhang2018}.

\emph{Many of these studies will benefit from a concerted effort between different groups to observe overlapping sky patches and to share data to maximize cross-correlation science output.}

\subsection{The Galaxy---Large-Scale-Structure Connection}
Mapping and characterizing the large-scale-structure (LSS) of filaments, sheets, knots and voids, commonly referred as the ``cosmic web", is essential to constraining the cosmology of our Universe and shedding light on several mechanisms driving galaxy formation and evolution.

Galaxies and galaxy clusters are preferentially formed along LSS filaments and knots, respectively, and are therefore biased tracers of the distribution of matter in the Universe.
LIM missions will map the emission in several spectral lines over large volumes. These maps can be used to reconstruct the underlying cosmic web and to perform several cosmological tests at different cosmic times. 
Moreover, the three-dimensional maps derived from LIM can be used to determine galaxy clustering and to correlate spectral-line luminosity of galaxies with their large-scale environment. Clustering studies are essential to access the probability of galaxy merger events, which are known to boost starburst activity and affect the overall galactic gas content. Also, galaxy positions relative to the LSS can affect their gas inflow rates. At high-z, where the EBL is very spatially inhomogeneous, a galaxy environment will likely correlate with the strength of the radiation field heating and ionizing its CGM. High-z clustering measurements are entirely out of reach for galaxy surveys, and any environmental constraint will be minimal. LIM thus uniquely addresses this important science goal.

\section{Reionization}

The epoch of reionization is an important yet virtually unexplored period in the history of the universe. During this period the first stars, accreting black holes and galaxies formed, emitted ultraviolet light, and gradually photoionized neutral hydrogen gas in their surroundings \cite{Loeb13}. The IGM during the first stages of reionization resembles a two-phase medium, composed of ``bubbles'' of highly ionized gas formed around luminous sources, while significantly neutral hydrogen regions remain intermixed. As reionization progresses, the bubbles begin to overlap until eventually, the IGM becomes highly ionized at a redshift close to $z\, =\, 6$.  

\emph{LIM can give a unique insight into the EoR by simultaneously mapping both the star-forming galaxies and quasars which produce ionizing photons, as well as the distribution of remaining neutral gas in the IGM.}
Some key, open questions which can be answered by LIM include: When did reionization occur? Stated otherwise, what is the ionized fraction of the IGM as a function of redshift? How large were the ionized regions at different stages of the reionization process? What was the relative contribution from the various sources of reionization? What were the thermal and chemical enrichment histories of the IGM?  How did the ionization of the IGM affect subsequent generations of galaxies? Answers to these questions will improve our understanding of the timing of early structure formation and the nature of the ionizing sources. 

\subsection{The Process of Reionization}

Multi-line intensity mapping directly probes the process of reionization as well as the sources that drive it.
Naturally, the most direct probe of the IGM during reionization is the redshifted 21-cm line; the first detections of 21-cm fluctuations from the EoR are anticipated in the next decade \citep{2015aska.confE...1K,DeBoer:2016tnn}. Currently, the best upper limit on this signal, by the LOFAR mission, is of $\Delta_{\rm 21}^2\, =\, (56\pm13\, {\rm mK})^2\, (1-\sigma)$ at ${\rm k\, =\, 0.053\, h{\rm cMpc}^{-1}}$ (and $\Delta_{\rm 21}^2\, <\, (300\pm13\, {\rm mK})^2\, (1-\sigma)$ at ${\rm k\, =\, 1.0\, h{\rm cMpc}^{-1}}$) \citep{2017ApJ...838...65P},  which is about two orders of magnitude above the expected signal. The LOFAR group will soon announce a tighter upper limit of $\Delta_{\rm 21}^2\, <\, (100\pm13\, {\rm mK})^2\, (1-\sigma)$ (Mertens et al. 2019, in prep) at ${\rm k\, =\, 1.0\, h{\rm cMpc}^{-1}}$, based on the analysis of 140h of data. Intensity maps of [CII], [OI], and [OIII] fine-structure lines, CO rotational transitions, and H-$\alpha$ emission, among others, in overlapping volumes, may be used to trace the galaxy distribution. Galaxy lines such as H$\alpha$, CO and [CII] can probe the rate of ionizing photons produced by star-forming galaxies and AGN as a function of time as well as the spatial distribution of these sources.
In conjunction, Lyman-$\alpha$ line maps, targeted by SPHEREx a NASA mission planed for launch in 2023 \cite{Dore:2018kgp} will trace both galaxies and the intergalactic gas. The proposed CDIM space mission \cite{Cooray2019} will also target the  Lyman-$\alpha$ line with a higher sensitivity and resolution and so with a much broader science case with respect to reionization studies.
The Lyman-$\alpha$ line can then be combined and cross-correlated with other probes, to put a series of constraints on the reionization process. This line is a particularly interesting EoR probe since shortly after being emitted most Lyman-$\alpha$ photons scatter off of neutral hydrogen in the interstellar, circumgalactic, and intergalactic media. 
Therefore during the EoR, the Lyman-$\alpha$ emission fluctuations are partly modulated due to the presence of ionized bubbles and can be used to probe the distribution of HI column densities in galaxies CGM.

Multi-line intensity mapping will thus probe a wide range of spatial scales and characterize the physical processes at play during reionization. Altogether, mapping both the galaxies responsible for emitting the ionizing radiation and the surrounding neutral gas in the IGM will dramatically improve our understanding of the fundamental interplay between the ionizing sources and the IGM \cite{Lidz:2008ry, Gong:2011mf, Lidz:2011dx}.

\section{The Cosmic Infrared Background}

\begin{figure}[h!]
\centering
\includegraphics[width=0.75\linewidth]{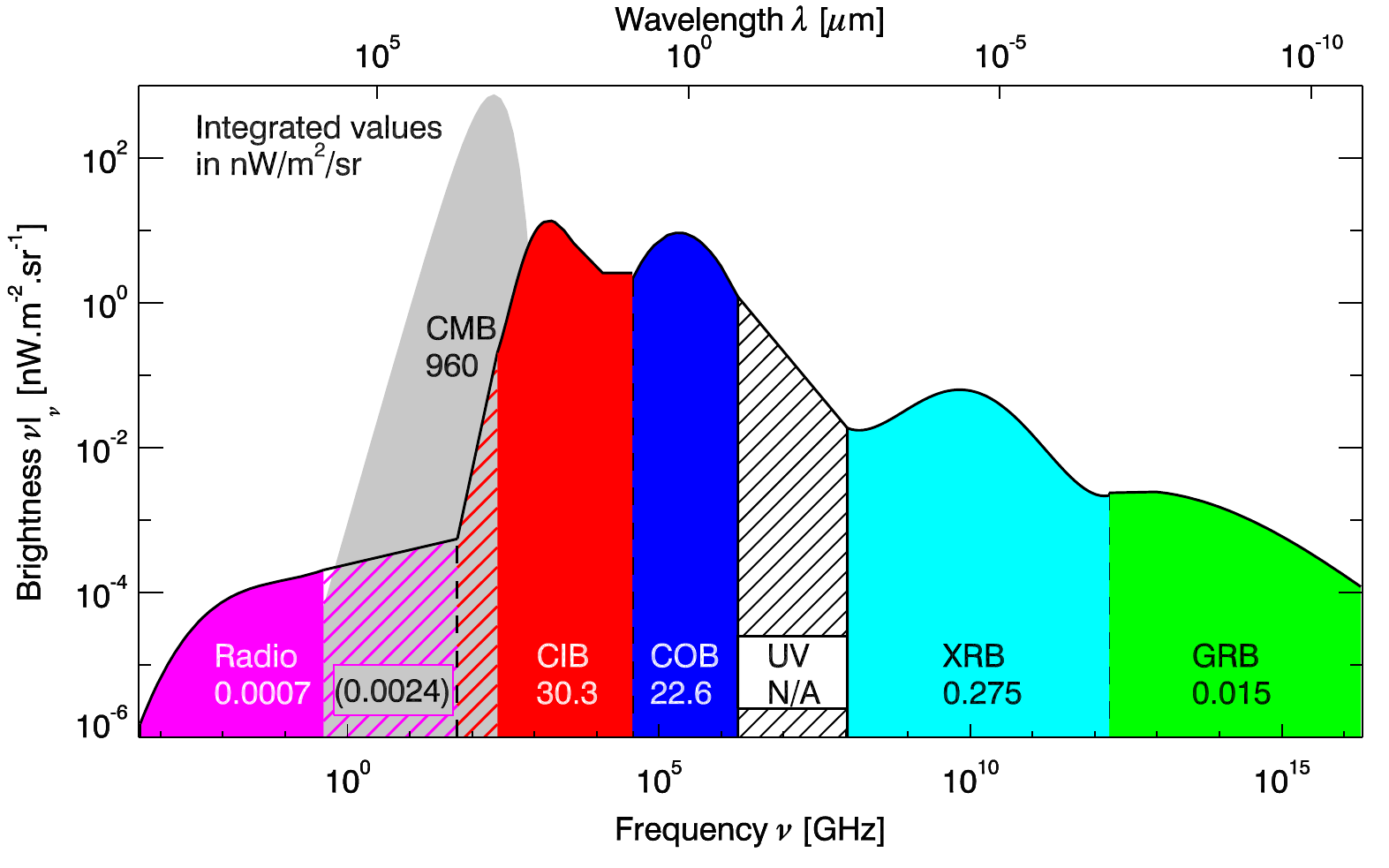}
\vspace{-0.1in}
\caption{Spectral energy distribution of the extragalactic background light from \cite{2011PhDT.......362B}. The cosmic infrared background in red is the second brightest background after the CMB (in grey).}
\label{fig:EBL}
\end{figure}

The extragalactic background light (EBL, see Fig.\,\ref{fig:EBL}) includes radiation from gamma-rays to radio and covers over 20 decades in wavelength \cite{Cooray2016EBL}.
Measuring and understanding its physical origin is a major goal of today's astrophysics/cosmology. The best-measured contributor to the EBL is the CMB which is measured to better than 1\%. The CIB is the second brightest background. It peaks around 160\,$\mu$m and extends from the mid-infrared to the millimeter domain. It is mainly emitted by dust in galaxies heated by the UV radiation from young freshly-formed stars. Contrary to the CMB, it is not emitted at the Epoch of Recombination but across cosmic times from the first galaxy to nowadays. The CIB is thus a valuable tracer of the evolution of galaxies and their host structures.

Absolute measurements of the CIB spectral energy distribution are challenging because of background and foreground subtraction (CMB, zodiacal light, cold galactic dust). Even the most recent analysis are only accurate at the level of $\sim$10\% combining FIRAS absolute photometry with the \textit{Planck} relative photometry \cite{2019ApJ...877...40O}.

To understand the build-up of the CIB, the contribution of the galaxies in the \textit{Spitzer} and \textit{Herschel} deep surveys has been estimated. Because of the confusion caused by the limited angular resolution of far-IR telescope, only a small fraction of the CIB can be resolved into individual bright sources. Stacking techniques have thus been developed to probe fainter sources and measure the CIB redshift distribution \cite{Bethermin2012,Viero2013b}. These studies showed that the entirety of the CIB can be explained by known galaxy populations (but the error bars are large) and that the CIB at longer wavelength is probing high-z emission (from the EoR).

Higher-precision measurement of the CIB would make it possible to differentiate between the total CIB and the contribution of the known galaxy population. This difference could come from astrophysical phenomenon as extended intergalactic dust emission (which would have strong implication for galaxy evolution models) or more exotic particle physics processes.

In complement to its absolute level, CIB anisotropies (and the cross-correlation with CMB lensing) provide critical information to identify the host halos of the dust-obscured star formation \cite{Maniyar2018}. These analysis relies on empirical models exploiting the fact that the redshift distribution of these anisotropies varies with wavelength \cite{Bethermin2013}. Unfortunately, the redshift distributions are broad and it is not possible to provide good constraints above z=3 due to degeneracies between spectral energy distributions (SEDs), clustering, and the star formation history. This last degeneracy could be broken in the future by improving the precision of the determination of the absolute CIB.

In principle, the CIB buildup can be determined by performing tomography of the CIB by cross-correlating it with spectroscopic galaxy samples \cite{Schmidt2015}. These type of measurements will constrain the average SED of the CIB at each redshift and will allow us to test if faint populations host particularly warm or cold dust. Finally, cross-correlating the CIB with galaxy lensing samples (e.g., Euclid) will probe directly how the star formation is distributed into the large-scale structures.

Cross-correlation analyses will allow dissecting the CIB at z$<$3. However, due to meaningful degeneracies between redshift slices, the best way to probe higher redshifts will be to use lines instead of the dust continuum. The CO and [CII] LIM are particularly promising because they are strongly correlated with star formation rate. And in particular, interesting constraints on the early production of dust in the Universe could be obtained by cross-correlating the CIB with EoR [CII] cubes, since they both trace IR emission.

Finally, as outlined in Section \ref{ssec:cii_mapping} LIM of [CII] will trace IR continuum emission in galaxies (a major contributor to the CIB), and as a result, this line intensity maps will characterize and pinpoint the location of the sources of the CIB anisotropies. Moreover, given the precise redshift information in [CII] maps, and the wide redshift range ($0\, <\, z\, <\, 8$) at which the assumed space mission will detect this line (with a high SNR), [CII] intensity maps will put precise and consistent constrains into the buildup of the CIB well into the EoR. 

\section{Protocluster identification with LIM}

The individual voxels in a line intensity map, contain the emission of a large number of galaxies belonging to several Halos.
However, the brightest voxels in a map will likely correspond to emission from (proto)clusters. At high-z, where the number density of bright sources is considerably smaller, the probability of the signal in a bright voxel being dominated by the emission of an individual protocluster is high. Therefore, LIM can be used to pinpoint the location, number density and clustering of these bright light sources which can later be followed up by galaxy surveys.

\emph{A LIM mission targeting emission at $z\, >\, 6$, will easily span large enough volumes to get a good statistical census of protoclusters and AGN.} Individually detecting/resolving protoclusters at these high redshifts requires a large SNR. Protoclusters are preferentially located in knots of the cosmic web. Therefore, in LIM, where the LSS is imaged, the location of bright voxels can be used to discriminate a protocluster from noise, with a signal to noise of a few. Overall, in LIM, protoclusters can be found with much modest sensitivity and spatial resolution than in other surveys. 
An LIM instrument with enough frequency coverage to detect two bright lines from the same structures at $z\, =\, 6$, such as CII and CO(7-6) (observed respectively at ${\rm 271\, GHz}$ and ${\rm 132\, GHz}$) can also be used to separate bright voxels in signal (protoclusters) or noise.

\section{Fundamental Cosmology}

Cosmology has only fairly recently transitioned from a theoretically-dominated discipline into an extremely active experimentally-driven field of research. What has become the concordance cosmological model, $\Lambda$CDM, is being repeatedly confronted with precision cosmology, which brings together massive amounts of data from various dedicated instruments that are rapidly growing in scale, and has done so quite successfully to date. 
Evermore accurate measurements of CMB anisotropies, complemented by input from increasingly deep galaxy surveys, have nailed down the parameters of the standard $\Lambda$CDM model to percent-level uncertainty~\cite{Aghanim:2018eyx}. 

However, several fundamental questions remain, touching on all three pillars of $\Lambda$CDM. While the inflationary paradigm is in good agreement with the data, we are lacking a more specific identification of the model of {\it inflation} that applies to our Universe. Meanwhile, we are largely still in the darkness with regards to the dark sector. The composition of {\it dark matter} and whether it interacts non-gravitationally with ordinary matter is unknown. As for {\it dark energy}, while in principle it may be explained by a cosmological constant (as the name of the model indicates), we cannot rule out more complicated descriptions (which indeed may be needed to explain the growing Hubble tension~\cite{Verde:2019ivm}).
From the observational standpoint, new ways to explore the uncharted high-redshift volume of the Universe must be developed.

{\em LIM is poised to revolutionize our understanding of these open questions.} The emitting species observed by LIM surveys are biased tracers of the underlying dark-matter density field, making them an excellent probe of large-scale structure.
Space-borne LIM experiments in particular can reveal invaluable new information by mapping significant parts of the sky over extended redshift epochs, and across a wide range of scales. The accessible number of  modes with intensity mapping of various lines stands to vastly outnumber that from the CMB anisotropies (which are already measured to cosmic variance uncertainty over most of the relevant range of scales) and from  galaxy number counts, which are limited to the nearby Universe (even future wide-field galaxy surveys will be able to probe only redshifts up to $z\lesssim3$).

\subsection{The cosmic expansion history}

One of the most promising applications of LIM to cosmology is filling the gap in the measured expansion history of the Universe. By measuring the anisotropic power spectrum of line-intensity maps at high redshift~\cite{Bernal:2019jdo}, taking advantage of the Alcock-Paczynski effect~\cite{Alcock:1979mp}, we can deduce the scale of baryon acoustic oscillations (BAO) at across cosmic times~\cite{Bernal:2019gfq}. The BAO scale provides a standard ruler which can be used to constrain the expansion rate. LIM BAO measurements at redshifts between $3<z<10$ can bridge over most of the gap between the local measurements of the expansion rate which rely on the local distance ladder~\cite{Riess:2016jrr}, and those inferred from measurements of CMB anisotropies at recombination~\cite{Aghanim:2018eyx}. 
This will be particularly crucial in light of the growing tension between the values stemming from these two different methods, which is now estimated to be as high as $5\sigma$~\cite{Riess:2019cxk,Wong:2019kwg}. 

\begin{figure}[h!]
\centering
\includegraphics[width=0.75\linewidth]{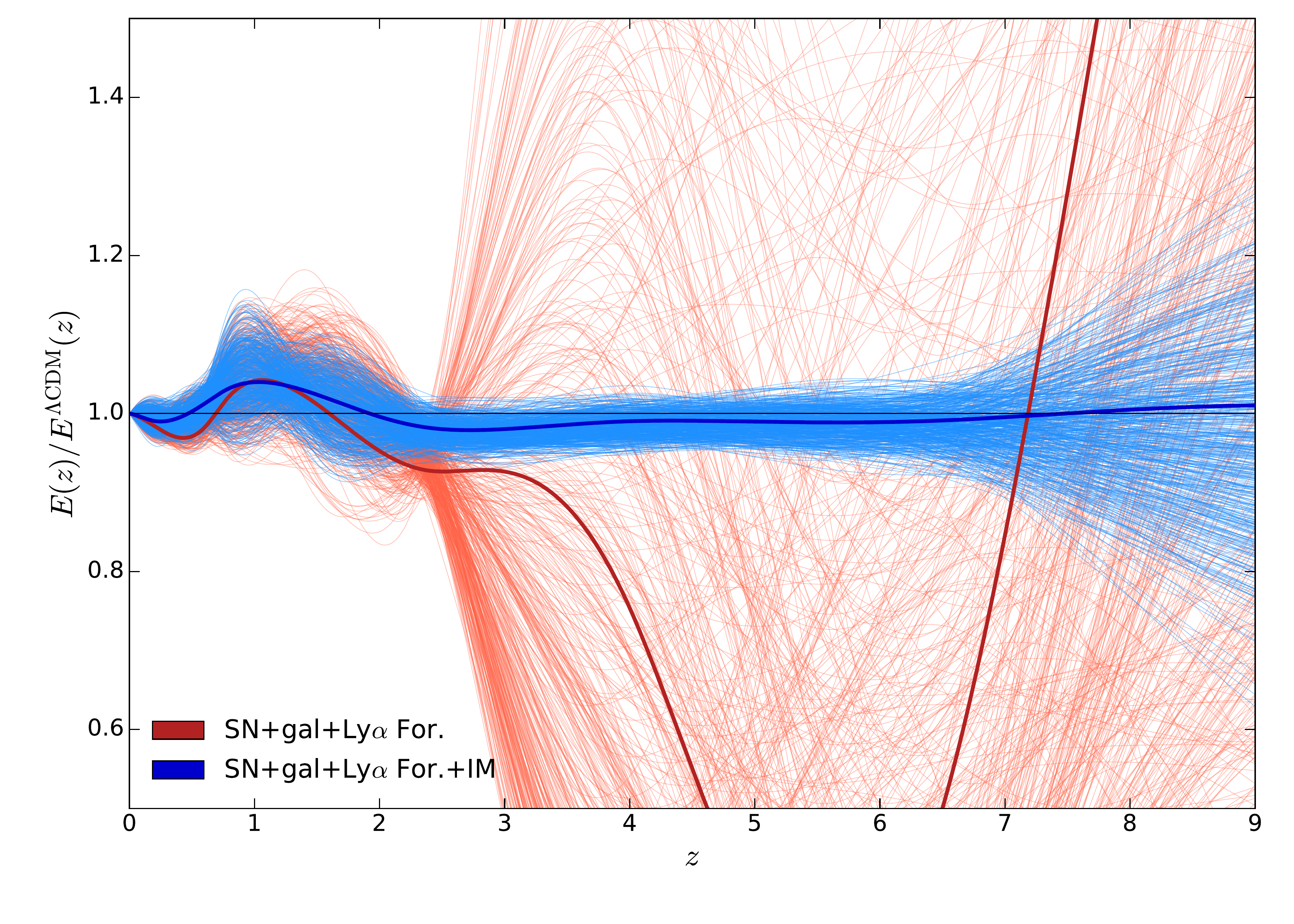}
\vspace{-0.1in}
\caption{Model-independent constraints on the shape of the cosmic expansion history, $E(z)=H(z)/H_0$ (normalised to $\Lambda$CDM). For details, see Refs.~\cite{Bernal:2016gxb,Bernal:2019gfq}. This illustration is calculated for a stage-3 LIM experiment measuring CO(1-0) over a 1000 deg$^2$ patch for 10,000 hours~\cite{Bernal:2019gfq}.}
\label{fig:expansion}
\end{figure}

In Figure~\ref{fig:expansion} we illustrate the power of CO LIM to constrain the cosmic expansion history at high redshifts, compared to using only Supernovae (SN), galaxy surveys and the Lyman-$\alpha$ forest. In comparison, the space mission proposed here will have superior signal-to-noise and will enable even stronger constraints (likely at the sub-percent level, provided that foreground contamination can be mitigated), using a combination of high-J CO lines and the CII line.

\subsection{Inflation and primordial non-Gaussianity}
Understanding the origin of structure in the universe is one of the key open questions in modern cosmology. Inflation is a leading paradigm of the early universe, in which quantum fluctuations during a phase of accelerated expansion set the seed of the structure in the universe. The primordial seeds evolve under the gravitational force, and give rise to the late-time structure we observe today. The simplest models of inflation (based on a single slow-rolling scalar degree of freedom, with a canonical kinetic term, and originating from a Bunch-Davies vacuum), predict a nearly Gaussian distribution of primordial fluctuations~\cite{Maldacena:2002vr}. Deviations from any of these assumptions can give rise to a distinctive non-Gaussian distribution, and hence distinctive shapes of higher-order statistics of primordial fluctuations. Commonly the primordial bispectrum and trispectrum are parameterized in terms of a shape and amplitude, and constraints on an amplitude of a given shape can discriminate between inflation models and the physical mechanism at play at the earliest stages of the universe and highest energy scales, inaccessible otherwise. 

The measurements of statistics of large-scale structure have the potential to significantly improve the current best constraints on primordial non-Gaussianity, set by measurements of CMB anisotropies by the Planck satellite \cite{Akrami:2019izv}, since there are many more modes in the 3D map of the universe vs. the 2D.  In addition to generating a non-zero contribution to the bispectrum and trispectrum of matter fluctuations and biased tracers, certain shapes of primordial non-Gaussianity, in particular the local shape, also leave an imprint on two-point statistics of biased tracers at the large scales. This is the so-called scale-dependant bias which has been used to constrain this type of non-Gaussianity from current galaxy surveys, and is considered the primary probe to constrain the local shape. 

Access to the largest scales is a critical requirement for using this signature as a probe of non-Gaussianity, which is why  LIM measurements have  great potential to provide exquisite constraining power due to their high-redshift/low-resolution nature, which allows to probe ultra large-scales, beyond the reach of traditional galaxy surveys. Since single-field models of inflation predict a very small amplitude of the local bispectrum, this configuration is considered a strong discriminant for multifield models. Based on using only this signature, studies have shown that LIM with emission lines can potentially approach the coveted target of $\sigma(f_{\rm NL})\!\sim\!1$~\cite{MoradinezhadDizgah:2018zrs,Fonseca2018}.  These constraints can further improve by considering also the line intensity bispectrum as well as combining the information of multiple lines in a multi-tracer analysis.       

We find that the proposed LIM space mission probing the CII line, in a 4-year survey with its high-resolution instrument, has the potential to constrain the amplitude of local-shape bispectrum with a 1-$\sigma$ uncertainty of $\sigma(f_{\rm NL}^{\rm loc}) \sim 0.75)$, after marginalizing over the parameters of $\Lambda$CDM cosmology. The expected constraining power depends on the theoretical model employed. The above constraint is obtained using the model outlined in section 1.2 for the CII intensity, accounting for the redshift-space distortion, as well as the Alcock-Paczynski effect.

\subsection{Probing $\Lambda$CDM}

With a large number of measured modes, LIM  can provide competitive constraints on the parameters of the standard $\Lambda$CDM model, augmenting and complementing other observables.

In table \ref{tab:cosmo_forecast} we show the forecasted constraints on cosmological parameters via CII intensity mapping with the described high-resolution instrument for $\Lambda{\rm CDM}$ cosmology, as well as two beyond-$\Lambda{\rm CDM}$ cosmologies, one with local non-Gaussianity of the local shape, and the other with effective number of neutrinos as a free parameter. Starting with the $\Lambda{\rm CDM}$ case, compared to the constraints from TT+TE+EE+lowE data, LIM with CII line can improve the constraints on Hubble parameter $h$, and the amplitude of primordial fluctuations $A_s$ by nearly an order of magnitude. The improvement on the tilt of the primordial spectrum $n_s$ is more modest, but there is still nearly a factor of 3 gain. When considering primordial non-Gaussianity, a target sensitivity of $f_{\rm NL}^{\rm loc} \sim 1$ is within reach, and in particular for the above survey spec, as mentioned above we forecast 1-$\sigma$ uncertainty of $\sigma(f_{\rm NL}^{\rm loc} = 0.75$, a factor of 9 better than current constraints from Planck. As for the number of relativistic degrees of freedom, the CII LIM mission can improve the Planck constraints by a factor of 5.        

\begin{table}[htbp!]
\centering
\begin{tabular}{c c c c }
\hline \hline
${\rm Parameters}$ 	& $\Lambda{\rm CDM}$ & $\Lambda{\rm CDM}+f_{\rm NL}^{\rm loc}$ &  $\Lambda{\rm CDM}+N_{\rm eff}$ \\ \hline
$\Omega_b$ &$1.4 \times 10^{-4}$ & $1.8 \times 10^{-4}$ & $4.0 \times 10^{-4}$ \\
$\Omega_c$ & $7.3 \times 10^{-4}$& $1.1 \times 10^{-3}$ & $1.2 \times 10^{-3}$ \\
$h$ & $7.2 \times 10^{-4}$ & $9.8 \times 10^{-4}$ &  $3.7 \times 10^{-3}$\\
$n_s$ &$1.3 \times 10^{-3}$ & $2.1 \times 10^{-3}$ &  $2.5 \times 10^{-3}$\\
$\ln(10^{10} A_s)$ &$2.8 \times 10^{-3}$ &$4.1 \times 10^{-3}$ &  $8.5 \times 10^{-3}$\\
\hline 
$f_{\rm NL}^{\rm loc}$ & - & 0.75 & -\\
$N_{\rm eff}$ & - & - &  0.071 \\
\hline
\end{tabular}
\caption{$1-\sigma$ uncertainties varying: only $\Lambda{\rm CDM}$ parameters; $\Lambda{\rm CDM}$ and amplitude of local bispectrum; and $\Lambda{\rm CDM}$ and the effective number of relativistic species.  Constraints for each of the cosmologies are obtained imposing Planck priors from plike(TT+TE+EE+lowE) chains.}
\label{tab:cosmo_forecast} 
\end{table}

Lastly, it is important to emphasize that LIM will also have crucial input regarding models of dark matter and of dark energy and modified gravity. For example, it has been shown that LIM can potentially improve the experimental sensitivity to radiative dark-matter decays or annihilation by up to ten orders of magnitude~\cite{Creque-Sarbinowski:2018ebl}, using cross-correlation of the spectral intensity maps with external traces of the mass distribution (such as CMB lensing). Meanwhile, the BAO measurements described above can probe models with a time-dependent equation of state for dark energy~\cite{Dinda2018} or modified gravity theories~\cite{Jain:2013wgs,Hall2013,Pourtsidou2016}.

\section{Required space mission to achieve science goals}

\subsection{Scientific requirements}

The main science goal of this white paper is to constrain the evolution of large scale structure in the Universe from present times to $z\sim8$ (middle stages of the EoR). We propose to meet this goal by imaging the cosmic web using the LIM technique to trace a series of galaxy lines in the sub-millimeter to FIR frequency bands. This mission will complement other third generation LIM missions, such as the SKA and CDIM, and therefore provide a comprehensive understanding of galaxy evolution and star formation over cosmic times. We propose to use [CII] LIM for the main cosmological breakthroughs and for the study of the high-z Universe. Therefore, mapping this line will be the main driver of the instrument requirements. Additional key science cases include measuring the SFRD from the peak of star formation to the EoR, constraining the CO SLED, and detecting a series of higher frequency FIR lines (such as [OIII] and [NII]) over a broad redshift range. 
Only a space-based mission can achieve the sensitivity and frequency coverage requirements to detect and image the structures emitting these lines. Current ground-based LIM missions have more modest science goals of making a preliminary line detections and testing the LIM associated technology and processing methodologies.

We now outline the science goals driving the instrument setup requirements:

\begin{enumerate}
    \item {\bf Probing the EoR with [CII] line tomography (Reionization, protoclusters, metal and dust enrichment at high-z):} Requires sensitive [CII] observations in the 6 to 8 redshift range plus removal of interloping lines in map space. Survey requirements: spectral coverage over the 200 - 400 GHz frequency range;  spectral resolution of R=300 (or higher); angular resolution of 1$^\prime$ (ideal) to 6$^\prime$ (minimum); sky coverage of 100 deg$^2$.
    \item {\bf  Cosmology with statistical tomography of [CII] and FIR lines:} Survey requirements: sky coverage of 1000 deg$^2$ (but ideally 10000 deg$^2$); spectral resolution of ${\rm R\, =\, 300}$ (minimum);  spectral coverage over the 200 - 2000 GHz frequency range (higher frequencies would make it possible to detect more FIR lines); sky survey sensitivity (minimum).
    \item {\bf  Tracing the SFRD history and CIB tomography with [CII]:} Survey requirements: spectral resolution of R=200-400;  spectral coverage over the 200 - 2000 GHz frequency range; sky coverage of minimum 100 deg$^2$; sky survey sensitivity.
    \item {\bf  Measuring the molecular gas content up to high-z with CO:} Also, includes constraining the CO SLED up to z$\sim8$ and probing the redshift evolution of galaxies ISM. Survey requirements: spectral coverage over the 50(minimum 100) - 400 GHz frequency range; spectral resolution of R=300 (minimum); sky coverage of 100 deg$^2$; deep survey sensitivity; angular resolution of 1$^\prime$ (ideal) to 6$^\prime$ (minimum). 
    \item {\bf  ISM and CGM of galaxies at z=2 with FIR lines plus AGN at high-z:} FIR lines can probe the ISM in galaxies and the CGM during the peak of star formation activity. The [OIII] line is a powerful tool to quantify the number density and distribution of AGN. Survey requirements: spectral resolution of R=200-400, spectral coverage over the 300 - 2000(3000) GHz frequency range; minimum sky coverage of ${\rm 100\, deg^2}$, deep survey sensitivity or higher.
    \item {\bf Cross-correlation and synergies with other surveys:} Survey requirements: wide frequency coverage, spectral resolution of R=200-400 (minimum); deep survey location should overlap with at least one HI 21-cm line field and preferably with several other galaxy/LIM surveys.
\end{enumerate}

The required frequency range from 100 to 2000\,GHz is only partially accessible from the ground, in main atmospheric windows around 150 and 240\,GHz. Observing at higher frequency from the ground is possible at selected frequencies, but very challenging if one wants to map significant areas of the sky. Only a space mission can provide the uninterrupted frequency coverage necessary to detect line emissions at all redshifts, and disambiguate between contributions from different lines at different redshifts.

\subsection{Mission profile}

The default space mission setup used in this study is outlined in Section \ref{Sec:1.2}. It assumes a full sky survey and a deep patch survey of 400 deg$^2$, a wide frequency span (100 - 2000 GHz) with spectral resolution R=300 and angular resolution of 1$^\prime$ at 300 GHz. 

\noindent {\bf Telescope size and operating temperature:} To achieve that angular resolution, a 3.5 meter size telescope (as already flown on the Herschel space mission), is required.
Photon noise from the sky and the environment is the ultimate limitation for observing the [CII], [NII] and [OIII] lines at high redshift, in the 200-600\,GHz frequency range. Reducing the part of the photon noise that comes from the instrument itself is crucial at those frequencies. To that effect, future space missions currently under study, such as the Origins Space Telescope or SPICA, make use of low-emissivity mirrors actively cooled to 4-6\,K. The gain in mapping speed, or equivalently in number of detectors required, is typically 3 orders of magnitude or more. To perform the above science programme, a cold telescope (10K or preferably less) is required.

\noindent {\bf Focal plane instrument:} Several technologies are available for measuring spectra at sub-mm wavelengths with a spectral resolution $\simeq 300$. Among those, on-chip filter-bank spectrometers, such as developed in Ref. \citep{Endo2019}, optimally collect photons over a large bandwidth with minimal weight and complexity. The baseline instrument considered here assumes 64 pixels near the sky photon noise limit, observing with a total light collecting efficiency of 30\% (due to losses in the optical filters, imperfect radiation coupling efficiency in the focal plane, and losses on the microstrip lines). To cover a wide frequency range, a few modules that cover about an octave each in frequency may be necessary. For achieving the required sensitivity, detectors must be cooled to sub-kelvin temperature. 

\noindent {\bf Readout electronics:} For observing the whole $50\, -\, 2000$ GHz frequency range at $R=300$ with 64 on-chip spectrometers, 
it is necessary to read-out $\simeq 70,000$ channels. A high rate of multiplexing is required to minimize the conductive thermal losses through the readout cables. Assuming 10 mW of power per detector, a total readout power of about 700W is required.

\noindent {\bf Orbit and operations:} An orbit around the L2 Sun-Earth Lagrange point, as for Herschel and Planck, offers the best observing conditions, with Sun, Earth, and Moon on the same side of the spacecraft. The operations require scanning large sky patches which are then connected to construct a full-sky map.

\noindent {\bf Data download:} Assuming sampling at about 10\,mHz, the scientific data rate is about 700 samples per second, or of the order of 10 kbit/s assuming two bytes per sample. The spacecraft can be pointed towards the Earth for data download.

\noindent {\bf Mission category and downscope options:}
The size and temperature of the telescope require an L-class mission for achieving these observations. However, the telescope and focal plane could be shared with other instruments with similar environment requirements, observing at nearby frequencies in a survey mode. 
An M-class mission with a smaller cold telescope (1-meter class, for a few arcminute angular resolution) could be envisaged as a down-scope option, but the telescope must remain cold for reaching the necessary sensitivity.

\section{Synergy with other Observations}
\subsection{Comparison and complementary with other LIM missions}

The LIM experimental landscape is flourishing, with numerous instruments already taking data, going online soon, having been approved, or going through the proposal stage~\cite{Bacon:2018dui,Koopmans:2015sua,Masui:2012zc,2013MNRAS.434.1239B,Newburgh:2016mwi,2013A&A...556A...2V,2014SPIE.9145E..22B,DeBoer:2016tnn,Pen:2008ut,Wu:2016vzu,2013PASA...30....7T,Parsons:2013dwa, Santos:2017qgq,Padmanabhan:2018yul,2015ApJ...814..140K,Li:2015gqa,Stacey:2018yqe,Crites:2014,2018A&A...609A.130L,2018AAS...23132804A,2016AAS...22742604B,Hill:2008mv,Gil-Marin:2014sta,Cooray:2016hro,2014arXiv1412.4872D,2018arXiv180909702T}. We outline some of the potential synergies between these experiments and our proposed instrument.

\begin{figure}[h!]
\centering
\includegraphics[width=\linewidth]{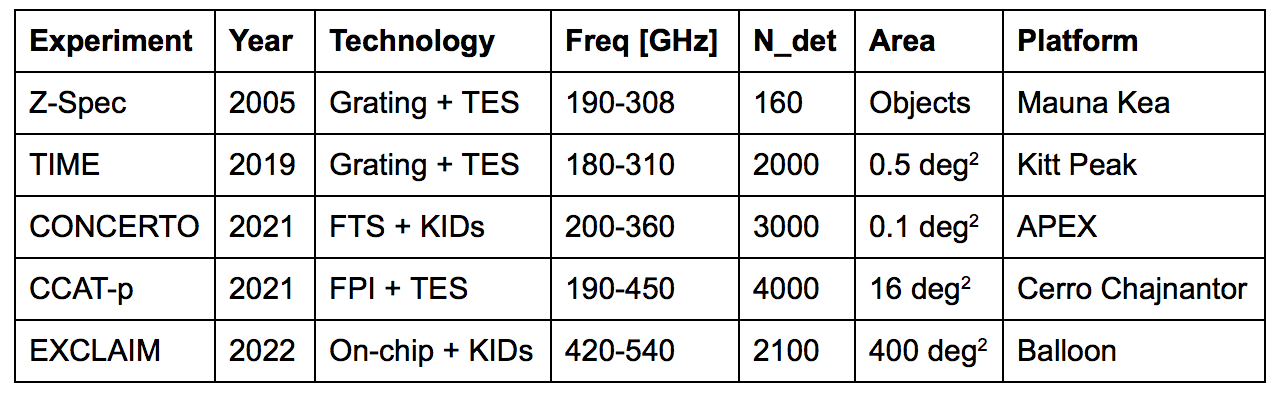}
\includegraphics[width=\linewidth]{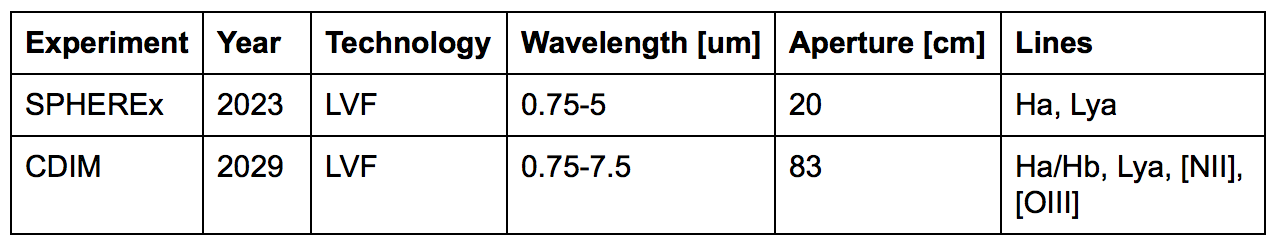}
\caption{Current LIM experimental roadmap for detection of CII at EoR redshifts and the high-J CO lines at medium redshifts ({\it top panel}) with ground and sub-orbital wideband mm-wave spectrometers; and for space missions targeting higher-frequency lines ({\it bottom panel}). In comparison to the instruments in the top panel, the proposed space mission will have much larger sky coverage and better sensitivity, allowing for efficient line separation, robust astrophysical constraints and meaningful cosmological measurements. Compared with the SPHEREx and CDIM  satellites it will target a complementary frequency range, bringing access to other lines, and thereby other physical processes and cosmological targets.}
\label{fig:IMExperiments}
\end{figure}

\noindent {\bf [CII] and high-J CO lines:} Top panel in Figure \ref{fig:IMExperiments} shows different LIM missions which will cover small sections of the frequency range probed by the proposed survey. TIME and CONCERTO will target [CII] at the EoR over small sky patches in order to achieve enough sensitivity for a statistical signal detection. Likely these missions will start by detecting CO high-J lines at low-z. In the context of these missions, new methods are being developed to address CO line contamination in CII maps at high-z. However, often these methods rely on simple assumptions about the astrophysics driving the CO lines and might not be as efficient with the real data. CCAT-p will target the same line on a much larger area which will help  separate the different lines. 
EXCLAIM will probe [CII] at $z\sim4$ over a very large sky area but with lower sensitivity then the proposed mission. Exclaim will target [CII] over a frequency range where this line dominates the observed signal in intensity but it will struggle with line contamination in power spectra analysis (due to projection effects foreground lines will show up with a high amplitude at the larger scales which are less affected by shot-noise). In  case of a detection, EXCLAIM can improve upon current constraints on dusty star formation during the epoch of stellar buildup.  
The proposed space mission has a big advantage over current missions since its line-intensity maps can be separated to the different interloping lines in map space. The line separation strategy will rely mostly on assuming that the different line signals are spatially correlated with the large scale structure. Therefore, the complex astrophysics driving the line strengths and specifically, the overall uncertainty of the CO emission will not have a meaningful effect on the line separation prospects of this mission. 

\noindent {\bf LIM of H-$\alpha$ and Lyman-$\alpha$ lines:} SPHEREx and CDIM will probe high frequency lines, by covering a broad frequency range over the UV and optical bands. The target lines will be emitted over a wide redshift range extending from $z<1$ to $z=6-10$. CDIM is the most sensitive mission of the two and in particular it has a strong science case for probing the EoR by constraining the intrinsic ionizing photon budget and the escape of ionizing photons with the H-$\alpha$, Lyman-$\alpha$ lines. These lines probe instantaneous star formation and their strength strongly correlates with galaxies ionizing photons emissivity. However, the H-$\alpha$ and Lyman-$\alpha$ lines suffer from dust extinction which needs to be corrected for. The proposed [CII] LIM mission would yield strong constraints on the dust extinction suffered by UV lines. SPHEREx will perform statistical tomography of these lines while CDIM proposes to image the emitting structures. Due to projection effects in statistical tomography (power spectra analysis), at scales close to $0.1\, Mpc$ the contamination by foreground lines (such as H-$\alpha$ in Ly-$\alpha$ maps will be amplified. CDIM maps can be analysed in map space where it is easier to separate the different lines. Still the proposed mission has the potential to provide the cleanest detection of a line emission during the EoR, namely the [CII] 158 $\mu$m line. This is due to a combination of a strong [CII] signal compared to that of its interloping lines and due to most of the multiple CO lines being strongly correlated.   

\noindent {\bf LIM of the CO(1-0) line:} The COMAP LIM mission will target the 26 to 34 frequency range and map CO(1-0) at $z\sim3$ in 4 patches of 2.5 deg$^2$. Moreover, this instrument will cover the CO(2-1) line at $z\sim6$ even if with low sensitivity to precisely constrain this signal.  The proposed survey will complement COMAP constraints by measuring high-J CO lines at the same redshift range and CO(1-0) at lower redshift. Cross-correlation between the proposed mission and COMAP could yield constrains on CO(2-1) line emission during the EoR ($z=6$).

\noindent{\bf LIM of Far-IR lines:} The OST \cite{2018Battersby} is a mission concept for a Far-IR space mission proposed to NASA. OST will operate in the wavelength range of 2.8 to 588 $\mu m$ ($\sim$ 550 GHz to 107 THz) with a sensitivity 1000 times higher than its predecessors. This instrument will partly overlap in frequency with the here proposed space mission and so it will also detect and characterize the FIR lines. However, OST will only probe [CII] up to $z\sim2.7$ and the most interesting constraints on the SFRD, EoR and cosmology require going higher in redshift. Nonetheless, OST can use LIM of a series of lines to probe how galaxies make metals, dust, and organic molecules and complement the constraints on the SFRD derived the [CII]/CO LIM at high-z.

\subsection{Synergies between LIM missions and other high-z probes}

LIM observations will complement, confirm and provide wider context for a series of other missions, spanning many different observables and various existing and planned instruments.

\noindent {\bf Deep galaxy observations:}
With its access to the cumulative signal from faint galaxy populations, LIM surveys will make it possible to determine the universality of various properties measured only for small numbers of bright galaxies at high redshift by instruments such as ALMA, the ngVLA and JWST. 
For example, it can test whether the source of ionizing photons at the epoch of reionization is mostly limited to low-luminosity galaxies (as suggested by current measurements~\cite{Robertson:2015uda}).
Moreover, LIM will enable clustering studies and relate line emission fluctuations with large scale structure fluctuations and growth. Also, LIM will enable cosmology studies at high-z which are beyond the capabilities of galaxy surveys, such as measuring baryon acoustic oscillations (BAOs) above z=3.

\noindent {\bf Large galaxy surveys:}
LIM measurements over large patches of sky and a wide range of redshifts will provide natural datasets for cross-correlation with galaxy catalogs from upcoming wide surveys such as WFIRST and LSST. 
This will enable a wide range of uses, from multi-tracer cosmological analyses to beat down cosmic variance, through removal of uncorrelated foregrounds such as our Milky-Way Galaxy, improving photometric estimations via the clustering-based redshift method~\cite{Menard:2013aaa,Cunnington2019}, to testing anomalies such as the strong emission attributed to the diffuse IGM~\cite{Croft:2018rwv}.

\noindent {\bf Lyman-$\alpha$ forest:} 
LIM can be combined with Lyman-$\alpha$ forest measurements from experiments such as 
  DESI~\cite{Aghamousa:2016zmz}, WEAVE~\cite{Pieri:2016jwo} and  CLAMATO~\cite{2018ApJS..237...31L} to help isolate and  constrain the impact of Lyman-$\alpha$ emitters on their environment, as well as to improve BAO measurements~\cite{Croft:2018rwv}. 
  A particularly interesting prospect is to combine LIM of a CO line by a LIM mission with tomographic Lyman-$\alpha$ forest from the CLAMATO survey. This could potentially yield important constraints on the neutral and molecular gas evolution in galaxies, as the ratio between gas phases is set by a complex tradeoff between a range of poorly constrained parameters such as metallicity, the ISM/CGM density distribution, stellar and AGN feedback, ISM turbulence, strength of the local and EBL radiation fields, etc.

\noindent {\bf CMB:} 
Correlations between LIM, which can map the cold gas distribution (as explained in Section 2 above), and CMB maps can be used to retrieve redshift information for secondary CMB anisotropies (e.g.\ lensing, or tracers of hot gas such as thermal and kinetic Sunyaev-Zel'dovich anisotropies~\cite{Ballardini:2018cho}), and to identify and characterize the galaxies/halos sourcing them.

\noindent {\bf Foreground Rejection:} 
LIM generically suffers from the problem of disentangling line interlopers from the target signal \cite{Kovetz:2017agg}. While cross-correlation between two lines observed by the same mission (a strong advantage of the space mission proposed here) or cross-correlations with other tracers of large-scale structure avoids foregrounds (the two lines will have uncorrelated foregrounds) and can improve the robustness of the two observations. 

\noindent {\bf Discovery space:}
The brightest voxels in LIM maps will pinpoint the brightest regions in space which can be followed up by galaxy surveys or dedicated observations with other telescopes. These regions are likely to contain some of the most extreme objects in our universe.
 LIM may also lead to the discovery or improved  characterization of new phenomena (the CHIME detection of fast radio bursts at frequencies $\gtrsim400\,{\rm MHz}$~\cite{Amiri:2019qbv} is a recent example).

\section{Summary}
Line-intensity mapping is a promising new observational technique which takes advantage of existing technology, developed for traditional galaxy surveys, to probe spectral line emission from galaxies and the IGM which is too faint or extended to be detectable by other surveys. 

The first generation of LIM missions has achieved initial detections at relatively low redshifts in 21 cm ($z\!\sim\! 0.8$) \cite{Masui13,Anderson2018}, [CII] ($z\! \sim\! 2.5$) \cite{Pullen:2017ogs,Yang:2019eoj}, and Lyman-$\alpha$ ($z\! \sim\! 3$) \cite{Croft:2015nna,Croft:2018rwv}, through cross-correlations with traditional galaxy or quasar surveys, as well as a CO auto-spectrum detection ($z \sim 3$) in the shot-noise regime \cite{Keating2016}. Over the next few years the COMAP mission will hopefully have secured a detection of the CO(1-0) line auto-spectrum at $z\sim3$. These should be the first of a series of auto power spectrum detections by second generation LIM missions throughout the coming decade. Namely, TIME and CONCERTO should detect CO high-J lines at low-z and [CII] at $z\sim6$. 

A number of efforts are already underway to improve LIM technology and push the statistical significance of these detections, as well as to increase the survey coverage in both redshift and volume. 
The next (third) generation of LIM missions promises to go beyond the level of mere detection to precise statistical tomography and imaging of the emitting structures. Detailed characterization of the LIM signals will yield competitive cosmology measurements as well as unbiased astrophysical constraints well beyond the capability of galaxy surveys. This next generation of LIM missions are still in the proposal phase and include the SKA2 (radio), CDIM (optical to UV band), the OST (Far-IR) and the sub-mm to FIR mission here proposed. The latter makes a nice bridge between the frequency ranges covered by COMAP and OST.

Preliminary estimates outlined in this paper show that a space mission covering the 100-2000 ${\rm GHz}$ frequency range could achieve sensitive measurements of [CII] during the EoR (SNR$>$10), low-J CO lines with SNR$\sim$50 (up to J=4-3), and of order $\sim10$ for high-J lines).  Galaxy properties derived from these lines will be well complemented with a few diagnostic FIR lines such as [OIII] (probing AGN) and [NII], detectable with  SNR$\lesssim$10 with the proposed deep survey. 

This sensitivity will make it feasible to accurately separate the different lines, opening a window to generating precise constraints of astrophysics and cosmological unknowns. For example, the high precision of the [CII]/CO measurements can be used to place tight constraints on the dust-enshrouded SFRD which is significantly beyond the capabilities of any existing or planned ground survey. Also, combined ratios of multiple CO lines and [CII] from the same structures, as well as precise clustering information obtained with LIM measurements, will make it possible to bring a better understanding of the relative importance of the different mechanisms driving star formation, such as molecular gas content and star formation efficiency.

[CII] intensity maps, obtained with the proposed space mission, will put strong constrains on the buildup of the CIB and identify and characterize the sources of CIB anisotropies.

Meanwhile, clean CO/[CII] intensity maps over large volumes of the observable Universe will access enough modes to yield competitive constraints on LCDM parameters and efficiently probe extensions to the simplest model such as primordial non-Gaussianity (reaching below the $\sigma(f_{NL})<1$ threshold), the number of effective relativistic degrees of freedom and the sum of neutrino masses. In addition, LIM BAO measurements can uniquely constrain the cosmic expansion history over redshifts $3<z\lesssim9$ (weighing-in on the growing Hubble tension by filling the gap between local and CMB determinations of the local expansion rate), and probe exotic models of dark matter, dark energy and modified gravity.

In conjunction with  upcoming surveys, it will be important to refine theoretical modeling efforts. New multi-scale simulation models are required to best capture the enormous range in spatial scale relevant for LIM observations, which involve the interstellar media of individual dwarf galaxies out to $\sim\!{\rm Gpc}$ cosmological length scales \cite{Villaescusa-Navarro2018,Stein2019}.  An essential step in the preparation for the post-processing and interpretation of observational maps is to improve the modeling of the target spectral lines and quantify their uncertainties given both existing and future constrains on galaxy properties and the underlying physics driving the line emission.  This will make it possible to predict/probe the link between line luminosity and the underlying astrophysical quantities and processes that we aim to constrain. Moreover, improving upon existing models will test the efficiency of current foreground removal methods and guide future strategies. Furthermore, cross-correlations and synergies between different surveys can only be understood if the mechanisms relating the line signals are taken into account.

To close, LIM is uniquely poised to address a broad range of science goals, from the history of star formation and galaxy evolution, through the details of the Epoch of Reionization, to critical questions in fundamental cosmology. 
This motivates an active research program over the coming decade, including continued investments in multiple line-intensity mapping experiments to span overlapping cosmological volumes, along with a strong support to increase simulation and modeling efforts. Thorough modeling of these lines signals adapted to each mission setup is essential on all mission phases, from instrument planning, to survey strategy, post-processing pipeline and interpretation of the observed signals in an astrophysics and cosmology context. {\em Above all, as this document describes, space is the ultimate frontier for line-intensity mapping. A satellite mission to target the CO and CII lines over a wide frequency range will  unlock the potential of this method to bring about game-changing results for astrophysics and cosmology.}

\newpage
\bibliographystyle{abbrv}
\bibliography{ScienceReport}

\newpage

\end{document}